\documentclass{article}
\usepackage{natbib}
\usepackage[margin=1in]{geometry}
\usepackage{mathrsfs}
\usepackage{amsmath,amsfonts,amssymb,amsthm,bm}
\usepackage{graphicx}
\usepackage{multirow}
\usepackage{array}
\usepackage{fixltx2e}
\usepackage{authblk}
\usepackage{listings}
\usepackage{hyperref}
\usepackage{color}

\usepackage[parfill]{parskip} 
\usepackage{pslatex} 
\newcommand{\pderiv}[2]{\frac{\partial #1}{\partial #2}}

\begin{document}
\title{\textsc{chrotran}: A mathematical and computational model for in situ heavy metal remediation in heterogeneous aquifers\footnote{Los Alamos National Laboratory technical report: \textbf{LA-UR-16-29041}}}

\author[1]{Scott K. Hansen\footnote{Corresponding author. skh3@lanl.gov}}
\author[1]{Sachin Pandey}
\author[1]{Satish Karra}
\author[1]{Velimir V. Vesselinov}
\affil[1]{Computational Earth Science Group (EES-16), Los Alamos National Laboratory}
\maketitle

\begin{abstract}
Groundwater contamination by heavy metals is a critical environmental problem for which in situ remediation is frequently the only viable treatment option. For such interventions, a three-dimensional reactive transport model of relevant biogeochemical processes is invaluable. To this end, we developed a model, \textsc{chrotran}, for in situ treatment, which includes full dynamics for five species: a heavy metal to be remediated, an electron donor, biomass, a nontoxic conservative bio-inhibitor, and a biocide. Direct abiotic reduction by donor-metal interaction as well as donor-driven biomass growth and bio-reduction are modeled, along with crucial processes such as donor sorption, bio-fouling and biomass death. Our software implementation handles heterogeneous flow fields, arbitrarily many chemical species and amendment injection points, and features full coupling between flow and reactive transport. We describe installation and usage and present two example simulations demonstrating its unique capabilities. One simulation suggests an unorthodox approach to remediation of Cr(VI) contamination.
\end{abstract}

\textbf{Keywords}: in situ remediation, abiotic reduction, bioremediation model, subsurface flow, reactive transport
\clearpage

\section{Introduction}
Heavy metals, including chromium, arsenic, copper, nickel, selenium, technetium, uranium, and zinc, are widespread and hazardous subsurface contaminants in groundwater aquifers \citep{Appelo2004,Tchounwou2012}.
For many heavy metals, their most stable oxidation state is often the most toxic \citep{Duruibe2007,Hashim2011}, 
and this oxidation state is typically the highest that occurs under near-surface conditions.
Additionally, the chemical reduction of certain metals is known to reduce their mobility \citep{violante2010}.
This has inspired efforts to manipulate in situ conditions to stimulate microbial growth and achieve biologically mediated metals reduction.
This technique has been demonstrated, at least in some settings, for chromium, uranium and selenium \citep{Lovley1993,Lovley1995}, nickel \citep{Zhan2012a}, technetium \citep{Istok2004}, and copper \citep{Andreazza2010}, and has been noted as a viable bioremediation technique by recent critical reviews \citep{Hashim2011,Wu2010}.
Bioprecipitation, a process by which microbiological exudates react with metals to produce an insoluble compound, has been widely observed \citep{malik2004,van2006,Radhika2006} and has been noted by \cite{Wu2010} as a remediation method.
Bio-stimulants have also recently been shown to effectively reduce chromium through abiotic redox pathways \citep{Chen2015,Hansen2016}, and after fermentation for other metals \citep{Hashim2011}.
Naturally, designing a remedial intervention using one of this family of techniques benefits greatly from the use of a multi-dimensional/multi-component numerical model of groundwater flow, contaminant transport, and biogeochemical processes to evaluate different remediation strategies under varying field conditions.
The model should be capable of capturing the transport behavior of electron donors, biomass, and other species, dominant biogeochemical reactions, and how these processes influence and are influenced by subsurface flow.

Although the development on in situ bioreactive transport models goes back to at least the 1980s, the literature is not vast.
Early work focused on in situ bioremediation of toxic organic compounds through oxidation.
A thorough mathematical and 2D numerical study representative of this approach is due to \cite{Chiang1991}, who presented a three-equation model involving a mobile electron donor (assumed to be the contaminant), mobile dissolved oxygen, and immobile biomass.
The contaminant was assumed to be consumed only in the microbial growth reaction, which was linear in biomass, Monod \citep{monod1949growth} in electron donor, and Monod in electron acceptor.
\cite{Wheeler1992} subsequently extended a reactive model of this sort to three dimensions to simulate biodegradation of $\mathrm{CH_4}$.
\cite{Travis1993} presented a more complicated, unsaturated three-dimensional model, which introduced Monod dependence on nutrients, and the potential for two electron donors, with one inhibiting the other.
This approach was further elaborated upon in a study of TCE degradation \citep{Travis1998a} by accounting for living and dead microbes, microbial predators, and first-order kinetic sorption of all aqueous species (microbes were treated as mobile).
Another complex oxidation model was developed by \cite{Suk2000}, which explicitly modeled both mobile and immobile biomass, contained a decay network, and featured both anaerobic and aerobic oxidation, in competition.

The development of models for metal reduction is comparatively more recent.
For U(VI), field-scale modeling studies have been performed on bio-reduction under anaerobic conditions at the Old Rifle Site in Colorado \citep{Li2010,Li2011,yabusaki2011}.
These conceptions treat the contaminant as the sole electron acceptor, with an externally applied electron donor, and the implied equations have a similar form to those devised by \cite{Chiang1991}: linear in biomass, Monod in contaminant, and Monod in electron donor.
For clarity, this is expressed symbolically as:
\begin{equation}
\frac{\partial C}{\partial t} \propto \frac{\partial D}{\partial t} \propto B \frac{C}{K_{C} + C} \frac{D}{K_D + D},
\label{eq: mu li}
\end{equation}
where $C$, is the U(VI) concentration, $B$ is the biomass concentration, $D$ is the electron donor concentration, $K_{C}$ is the metal reduction Monod constant, and $K_D$ is the electron donor Monod constant.
$K_{C}$ and $K_{D}$ respectively represent the concentration of $C$ and $D$ at which the reaction rate is halved.
Recently, \cite{molins2015} have published a numerical study of a column experiment with multiple species, all of whose dynamics are of the above form, but including an extra chemical inhibition factor. 
The models of \cite{Li2010,Li2011} were implemented at field-scale in CrunchFlow \citep{Steefel2015}, using its capability to represent single and multiple Monod formulations.

Systems of governing reactive transport equations for enzymatic microbial Cr(VI) reduction have been presented by \cite{Alam2004}, and by \cite{Shashidhar2007}.
\cite{Shashidhar2007} described the Cr(VI) degradation reaction slightly differently from \cite{Li2011}:
\begin{eqnarray}
\frac{\partial C}{\partial t} \propto \frac{\partial D}{\partial t} \propto B \frac{K_{C'}}{K_{C'} + C} \frac{D}{K_D + D}.
\label{eq: mu alam}
\end{eqnarray}
$K_{C'}$ is the concentration of Cr(VI) at which the reaction rate is halved, which is similar to $K_C$.
However, although (\ref{eq: mu alam}) appears superficially similar to (\ref{eq: mu li}), the $C$ factor represents entirely different behavior: not as an energy source but rather as an inhibitor.
Interestingly, since the RHS of (\ref{eq: mu alam}) is a proxy for the biomass growth reaction, $C$ consumption is modeled as proportional to biomass growth, but the biomass growth rate is modeled as independent of $C$.
Biomass dynamics were governed by a growth term proportional to donor consumption and a first-order decay term, accounting for eventual biomass die-off.
Other authors \citep[e.g.,][]{Somasundaram2009} have used a similar approach.
\cite{Alam2004} presented a relatively complex model which included transport with both mobile and immobile biomass, and also included two enzymes (both created due to biomass growth, but one conserved, and one irreversibly consumed during bio-reduction).
Neglecting the irreversibly-consumed enzyme and the mobile-immobile behavior, this model shares its electron donor and biomass dynamics with the model of \cite{Shashidhar2007}.
It differs significantly from other models that we are aware of by treating the Cr(VI) degradation reaction in this model as an incidental enzymatic process, and is governed by the following Monod equation:
\begin{eqnarray}
\frac{\partial C}{\partial t} \propto B \frac{C}{K_{C} + C}.
\label{eq: cr monod}
\end{eqnarray}
There is strong experimental support for this approach \citep[e.g.,][]{Okeke2008}, and this is arguably more defensible in a real complex geochemical system in which there are multiple competing donors and receptors, and given that there is evidence for indirect reduction pathways, e.g., by metabolites \citep{Priester2006}.
All of the models of Cr(VI) bio-reduction, discussed above, appear to be one-dimensional only.

Our literature review did not reveal discussion of field-scale bio-reduction models for other heavy metal species.
It thus appears that the primary example of a bio-reduction model applicable to modeling a real-world remediation scheme is the CrunchFlow model of uranium treatment at the Rifle site, which was discussed above.
We set out to develop a new model, dubbed \textsc{chrotran}, which is optimized for modeling bioremediation of Cr(VI), but of sufficient generality that it might be used for bioremediation of other metals, or for abiotic reduction, with ease.
The key features of the model we developed are as follow:
\begin{description}
	\item[Direct abiotic reaction between electron donor and contaminant] Recent experimental results \citep{Chen2015,Hansen2016} have established a rapid direct redox reaction when molasses is used as an electron donor and Cr(VI) is the contaminant, rather than the bio-mediated reaction previously posited.
	It is thus crucial to include this behavior in a model aimed at remediation design.
	\item [Indirect Monod kinetics] On account of the evidence \citep{Wang1995,Okeke2008,Hansen2016} for modeling Cr(VI) degradation with (\ref{eq: cr monod}), we implemented this general formulation as opposed to one which ties all contaminant degradation to a single biomass growth equation. 
	\item [Bio-fouling / Bio-clogging] It is well known in practice that one of the problems afflicting bioremediation schemes is build-up of biological material near the amendment injection point.
	This reduces the hydraulic conductivity, interfering with amendment injection, and may rapidly consume any amendment that does manage to pass through it.
	The model thus contains feedback between local biomass concentration and flow parameters such as porosity and hydraulic conductivity.
	\item [Biomass crowding] Similarly, if biomass becomes overly dense, this causes cell stress, which reduces the rate of further growth.
	Since clogging is enabled, this behavior was added as well.
	\item [Modeling of amendment additives] To address clogging or to attempt to spread electron donors farther from the well before they are consumed, additional chemicals may be injected to reduce biomass concentrations, and their reactive transport behavior is incorporated.
	\item [Multiple donor consumption pathways] The best model of electron donor consumption by biomass may be proportional to biomass concentration or biomass growth, and the model can handle any such combination.
\end{description}
Building this functionality required custom programming beyond what is embedded in existing reactive transport codes \citep{Steefel2015}.
To accomplish our goal, we turned to \textsc{pflotran} \citep{lichtner2015pflotran}, which is open source, has a modular structure, and a ``reaction sandbox'' interface \citep{hammond2015pflotran} that allows derivative versions with custom reaction behavior to be developed and compiled. In this fashion, no changes to the flow and transport part of \textsc{pflotran} are needed.
We developed \textsc{chrotran} based on the existing \textsc{pflotran} code framework, taking advantage of the reaction sandbox interface to implement complex model features not included in basic microbial packages and leverage other aspects of \textsc{pflotran}, such as its high-performance computing capabilities.

In Section 2, we present the mathematical details of \textsc{chrotran} and justify some of the decisions underlying the model.
In Section 3, we discuss how to install and use the software.
In Section 4, we present two numerical studies which illustrate \textsc{chrotran} and also suggest an interesting conclusion regarding Cr(VI) remediation.
In Section 5, we briefly summarize what has been presented.

\section{Model description}
\label{sec: equations}
We consider flow and transport at aquifer scale.
Conceptually, the aquifer is modeled as saturated, with incompressible water moving in accordance with Darcy's law (\textsc{chrotran} can also simulate partially-saturated vadose-zone flow and transport).
Two transport processes are considered, namely, advection with the Darcy flow and Fickian dispersion.
Multiple reaction terms are then added in order to capture the complex chemical dynamics during remediation.
As the model is intended to be used for remedial design, every effort was made to simplify the formulation to use the smallest number of explanatory variables and parameters, and to keep the equations at a high level of abstraction, so they are not tied to one particular set of chemical species.

The following are the several species whose dynamics are captured by the system of reaction equations, each with their own symbols:
\begin{description}
	\item[Biomass, $B \: \mathrm{[mol\:L_b^{-1}]}$,] representing the concentration of all microbes and their associated extracellular material.
	The quantification of biomass as a ``molar'' rather than a mass concentration is unusual, and was done for two reasons: (i) to avoid hard-coding units in which biomass concentration is to be specified, and (ii) to simplify presentation of the model, so all governing equations have the same units. A mole of biomass should be understood as an equivalent mass: any quantity can be used, as long as one uses a consistent definition throughout the model. In the examples in this paper, we use the definition 1 mol $\equiv$ 1 g of biomass.
	\item[Aqueous contaminant, $C \: \mathrm{[mol\:L^{-1}]}$,] which we here assume is a heavy metal ion in its oxidized state, such as Cr(VI) or U(VI).
	\item[Electron donor,] which is part of the chemical amendment, and may be
	\begin{enumerate}\setlength\itemsep{-0.1cm}
		\item immobile, represented by $D_i \: \mathrm{[mol\:L_b^{-1}]}$, or
		\item mobile, represented by $D_m \: \mathrm{[mol\:L^{-1}]}$,
	\end{enumerate}
	with exchange of mass between the two states.
	\item[Nonlethal biomass-growth inhibitor, $I \: \mathrm{[mol\:L^{-1}]}$,] such as ethanol, which is modeled as a conservative species but acts to slow microbial growth. 
	\item[Biocide, $X \: \mathrm{[mol\:L^{-1}]}$,] which reacts directly with biomass and is consumed.
\end{description}
For convenience, we also define a total species aqueous concentration of the electron donor, $D$, according to the formula $D = \frac{D_i}{\theta(\bm{x},t)} + D_m \: \mathrm{[mol\:L^{-1}]}$, where $\theta(\bm{x},t)$ [-] is the current porosity at $\bm{x}$.
For simplicity, we assume that both the mobile and immobile donor participate equally in all reactions.

\subsection{Flow and transport}
\subsubsection{Groundwater flow equations}
Flow may be modeled using the balance of water mass given by 
\begin{equation}
\frac{d}{dt}\left(\rho_w \theta\right) + \nabla\cdot\bm{q} = q_M(\bm{x},t),
\end{equation}
with the water mass fluxes related to head via Darcy's law:
\begin{equation}
\bm{q}=-\nabla\left(\rho_w K(\bm{x,t})h(\bm{x},t)\right),
\end{equation}
where $K(\bm{x}) \ \mathrm{[m/s]}$ is the local hydraulic conductivity and $h(\bm{x},t) \ \mathrm{[m]}$ is the local hydraulic head, $\theta$ is the porosity and 
$q_M(\bm{x}) \ \mathrm{[kg\ m^{-3}\ s^{-1}]}$ is the local mass injection rate into the system and $\rho_w \ \mathrm{[kg\ m^{-3}]}$ is the density of water.
We note that, since \textsc{chrotran} is built on top of \textsc{pflotran}, it inherits all of \textsc{pflotran}'s groundwater flow modeling capabilities.
This includes the ability to consider unsaturated and otherwise multiphase flow conditions, which are out-of-scope for the present discussion.
Please see the \textsc{pflotran} user manual \citep{lichtner2015pflotran} for details on its complete capabilities.

The hydraulic conductivity is continually updated in accord with the relation
\begin{eqnarray}
K(\bm{x},t)=K(\bm{x},0)\frac{\theta(\bm{x},t)}{\theta_0},
\end{eqnarray}
where $\theta_0$ [-] is the spatially-uniform initial porosity, and $\theta(\bm{x},t)$ is calculated according to
\begin{equation}
\theta(\bm{x},t) = \theta_0 - \frac{B(\bm{x},t)}{\rho_B},
\label{eq: biovfupdate}
\end{equation}
where $\rho_B \ \mathrm{[mol\ L^{-1}]}$ is the intrinsic biomass density.
(Note that, using our proposed definition of 1 mol of biomass as 1 g of biomass, 1 $\mathrm{mol\ L^{-1}} =$ 1 $\mathrm{kg\ m^{-3}}$.)

\subsubsection{Advective-dispersive transport operator}

We define $\mathscr{T}\{ \cdot \}$ to be an advective-dispersive transport operator, which characterizes the hydrodynamic effects on solute transport.
For $c$, the concentration of an arbitrary mobile species,
\begin{equation}
\mathscr{T}\{c\} \equiv -\bm{q}\cdot\nabla c + \nabla\cdot(\theta\bm{D}(\bm{q})\nabla c), 
\end{equation}
where $\bm{D}$ is a dispersion tensor that depends on the longitudinal and transverse dispersivities, molecular diffusion as well as the Darcy flux. For the work in this paper, we will only consider molecular diffusion and thereby we set $\bm{D} = D_m \bm{I}.$
Note that, while this is not shown explicitly for compactness, all symbols in this equation are functions of $\bm{x}$ and $t$.

\subsection{Biogeochemical reactions}

We define one governing equation for each species, mobile or immobile, as well as two equations defining reaction rate expressions for algebraic convenience.
The equations involve numerous parameters, whose symbols, units, and long-form name in the \textsc{chrotran} input file are summarized in Table \ref{tab: vartiable names}.
The parameter symbols follow a scheme in which the first letter encodes the physical interpretation of the parameter and the subscript specifies the governing equation in which they participate.
A symbol beginning with $\Gamma$ is a second-order mass action rate constant, with units $[\mathrm{L\ mol^{-1}\ s}]$.
A symbol beginning with $K$ is Monod or inhibition constant with units of concentration, $\mathrm{mol\ L_b^{-1}}$ or $\mathrm{mol\ L^{-1}}$, and represents the concentration at which a process rate becomes 50\% of its maximum rate, all other parameters being equal.
A symbol beginning with $\lambda$ has units of $\mathrm{s^{-1}}$ and is interpreted as a pure first-order reaction rate constant.
A symbol beginning with $S$ is dimensionless, and represents a stoichiometric relationship between a reaction rate and the consumption rate of a certain species.
Before presenting the equations, it is useful to review all of the chemical processes that are incorporated into the model:

\begin{description}
	\item[Abiotic reduction] This is an aqueous-phase bimolecular reaction between the electron donor, $D$, and the contaminant, $C$.
	It is modeled with a classical second-order mass action rate law. 

	\item[Bio-reduction] This represents the removal of the contaminant, $C$ by the biomass, $B$.
	The process is linear in $B$, and Monod in $C$, which is a common assumption for bio-mediated processes.
	Note that we are not assuming that reduction of $C$ is directly tied to any particular cell metabolic process.
	This form is sufficiently general that it can capture other bio-remediation processes besides bio-reduction of heavy metals.

	\item[Biocide reaction] This is an inter-phase bimolecular reaction between the biocide, $X$, and the biomass, $B$.
	It is modeled with a classical second-order mass action rate law, with the added condition that $B$ cannot fall below a specified minimum concentration $B_\mathrm{min}$.

	\item[Biomass growth] The core biomass growth reaction irreversibly consumes electron donor, $D$ to increase biomass, $B$.
	As a biologically catalyzed reaction, it is assumed to be linear in $B$ and Monod in $C$.
	Two inhibition effects are assumed: a biomass crowding term, tunable with exponent $\alpha$, attenuates growth rate as the biomass concentration rises.
	The nonlethal inhibitor concentration, $I$, also reduces the reaction rate as its concentration increases.

	\item[Mobile-immobile mass transfer (MIMT)] This is a process with first-order kinetics, which models sorptive retardation of the electron donor.

	\item[Natural decay] This is an empirical process reflecting the idea that, if left unstimulated, both the amount of living cells and the amount of extracellular material in the aquifer will ultimately return to their natural background level (i.e. $B_\mathrm{min}$).
	This is modeled as a first-order process.

	\item[Respiration] This represents consumption of the electron donor for purposes of life maintenance, unrelated to biomass growth.
	This is described by a first-order rate law which is proportional to biomass concentration, $B$. 
\end{description}

The explanations of the operative processes and of parameter interpretation above help the descriptions of factors and terms in the governing equations presented below.

\subsection{Reactive transport equations}
\label{sec: rt eqns}		
\subsubsection{Definitions of convenience reaction variables}
The biomass growth reaction is linear in biomass concentration, has a Monod dependency on electron donor, a tunable inhibition factor due to biomass crowding, and a classic inhibition factor describing the impact of the nonlethal growth inhibitor (as indicated by comment braces):

\begin{equation}
\mu_{B} = \lambda_{B_1}B
\overbrace{\frac{D}{K_D+D}}^{e^-\mathrm{donor}}
\underbrace{\left(\frac{K_B}{K_B+B}\right)^\alpha}_{\mathrm{crowding}}
\overbrace{\frac{K_I}{K_I+I}}^{\mathrm{inhibition}} \qquad \mathrm{\left[\frac{mol}{L_b\ s}\right]}.
\end{equation}

The direct, abiotic reduction reaction is represented by a classic, second-order mass action law:

\begin{equation}
\mu_{CD} = \Gamma_{CD} C D \qquad \mathrm{\left[\frac{mol}{L_b\ s}\right]}.
\end{equation}

\subsubsection{Partial differential equations for mobile chemical components}
The mobile components are all governed by the advection-dispersion operator, $\mathscr{T}$, defined previously, and also affected by extra terms implementing the chemical processes outlined earlier (as indicated by comment braces):

\begin{eqnarray}
\pderiv{\theta C}{t} &=& \mathscr{T}\{C\} -\overbrace{\lambda_{C}B\frac{C}{K_{C}+C}}^{\mathrm{bio-reduction}}-\underbrace{S_{C}\mu_{CD}}_{\mathrm{abiotic\;reduction}} \qquad \mathrm{\left[\frac{mol}{L_b\ s}\right]}, \\ [6pt]
\pderiv{\theta D_m}{t} &=& 
\mathscr{T}\{D_m\}-\overbrace{S_{D_1}\frac{D_m}{D}\mu_B}^{\mathrm{biomass\ growth}}-
\underbrace{\lambda_{D}\frac{D_m}{D}B}_{\mathrm{respiration}}-
\overbrace{S_{D_2} \mu_{CD} \frac{D_m}{D}}^{\mathrm{abiotic\; reduction}}-
\underbrace{\lambda_{D_i} \theta D_m+\lambda_{D_m}D_i}_{\mathrm{MIMT}} \qquad \mathrm{\left[\frac{mol}{L_b\ s}\right]},\\ [6pt]
\pderiv{\theta I}{t} &=& \mathscr{T}\{I\} \qquad \mathrm{\left[\frac{mol}{L_b\ s}\right]},\\ [6pt]
\pderiv{\theta X}{t} &=& \mathscr{T}\{X\}-\overbrace{\Gamma_{X}BX}^{\mathrm{biocide\; reaction}} \qquad \mathrm{\left[\frac{mol}{L_b\ s}\right]}.
\end{eqnarray}

\subsubsection{Partial differential equations for immobile chemical components}
The immobile component concentrations are affected only by the reactive processes outlined above:

\begin{eqnarray}
\pderiv{B}{t} &=& 
\overbrace{\mu_B}^{\mathrm{biomass\ growth}}-
\underbrace{\lambda_{B_2}(B-B_\mathrm{min})}_{\mathrm{natural\;decay}}-
\overbrace{\Gamma_{B}(B-B_\mathrm{min})X}^{\mathrm{biocide\;reaction}} \qquad \mathrm{\left[\frac{mol}{L_b\ s}\right]},\\ [6pt]
\pderiv{D_i}{t} &=& 
-\overbrace{S_{D_1} \frac{D_i}{D}\mu_B}^{\mathrm{biomass\ growth}}-
\underbrace{\lambda_{D}\frac{D_i}{D}B}_{\mathrm{respiration}}-
\overbrace{S_{D_2}\mu_{CD} \frac{D_i}{D}}^{\mathrm{abiotic\; reduction}}+
\underbrace{\lambda_{D_i} \theta D_m-\lambda_{D_m}D_i}_{\mathrm{MIMT}} \qquad \mathrm{\left[\frac{mol}{L_b\ s}\right]}.
\end{eqnarray}

\section{Software implementation}
\subsection{Software availability and installation}
\textsc{chrotran} currently is undergoing an open source licensing process, and will be available as soon as possible as freely downloadable Fortran 2003 source code. Until this process is complete, the software may be made available on a restricted basis by arrangement with the authors.

\textsc{chrotran} must be compiled using the GFortran complier (freely available as part of the GNU Compiler Collection).
It is based on the open-source \textsc{pflotran} code base, and the installation procedure is the essentially the same as that required to build \textsc{pflotran} from source, and \textsc{chrotran} requires all the libraries upon which \textsc{pflotran} depends, including PETSc \citep{balaypetsc} and others.
For installation of required libraries, the \textsc{pflotran} installation instructions\footnote{Available at \url{http://documentation.pflotran.org/user_guide/how_to/installation/installation.html}.} are applicable, except that \textsc{chrotran}, rather than \textsc{pflotran}, should be cloned from its repository\footnote{Not officially released yet; anonymous access can be facilitated through editor for review purposes.}
once all the dependencies have installed.
To build \textsc{chrotran} itself, navigate to \texttt{<path of cloned repository>/src/pflotran} and type \texttt{make chrotran}. (The \textsc{chrotran} executable will be called \texttt{chrotran}.)

\subsection{Specifying and running a simulation}
\begin{figure}
\includegraphics[trim={2.5cm 3.5cm 5cm 2.5cm},clip]{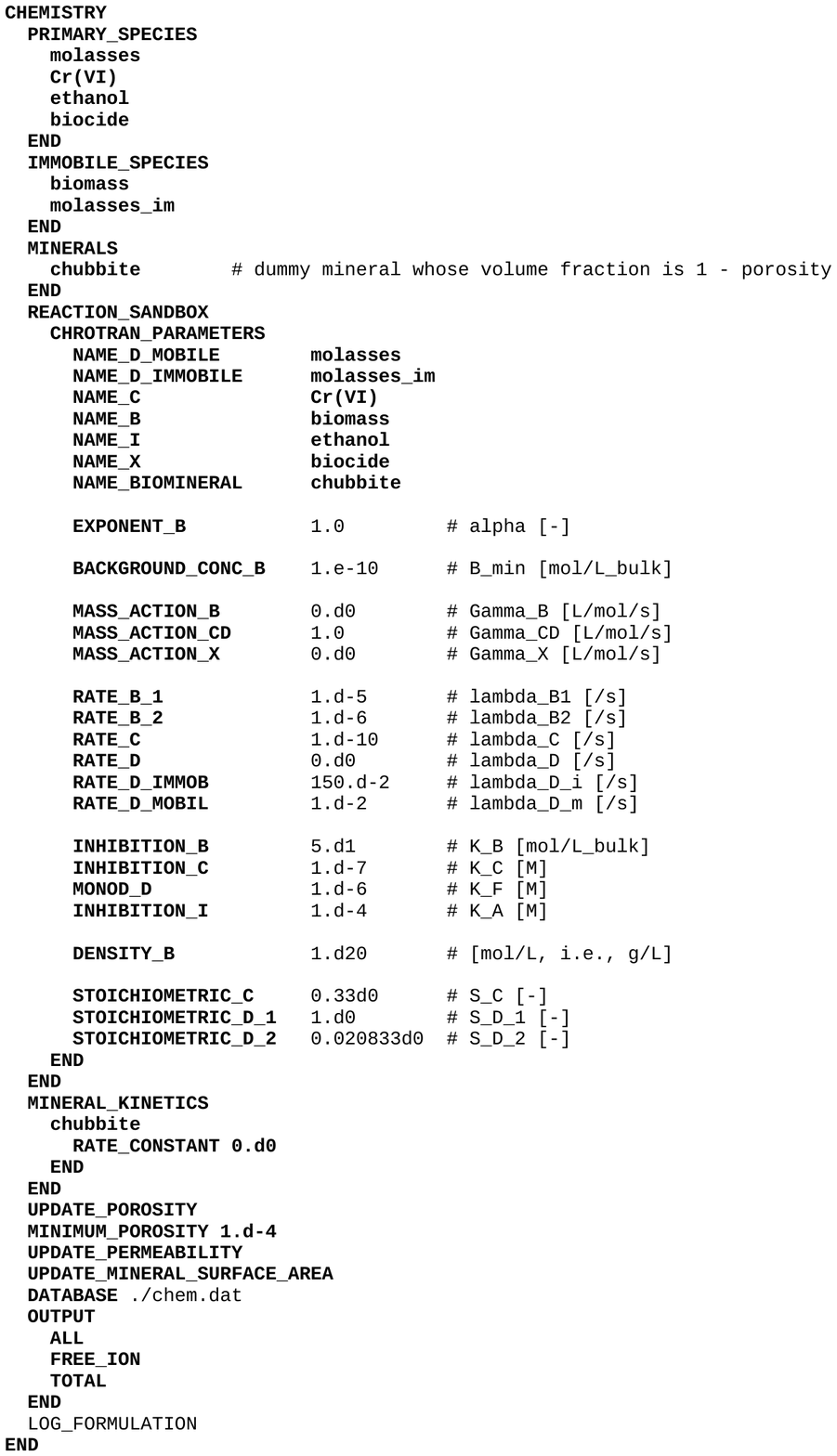}
\caption{Example \texttt{CHEMISTRY} card for \textsc{chrotran} input file.
Bold text should not be altered.
However, additional species may be added to the \texttt{PRIMARY\_SPECIES}, \texttt{IMMOBILE\_SPECIES}, \texttt{MINERALS}, and \texttt{MINERAL\_KINETICS} blocks, if desired.
Additional sandboxes can also be used in the \texttt{REACTION\_SANDBOX} block.}
\label{lst: chemistry}
\end{figure}

\begin{figure}
\includegraphics[trim={2.5cm 13cm 10cm 2.5cm},clip]{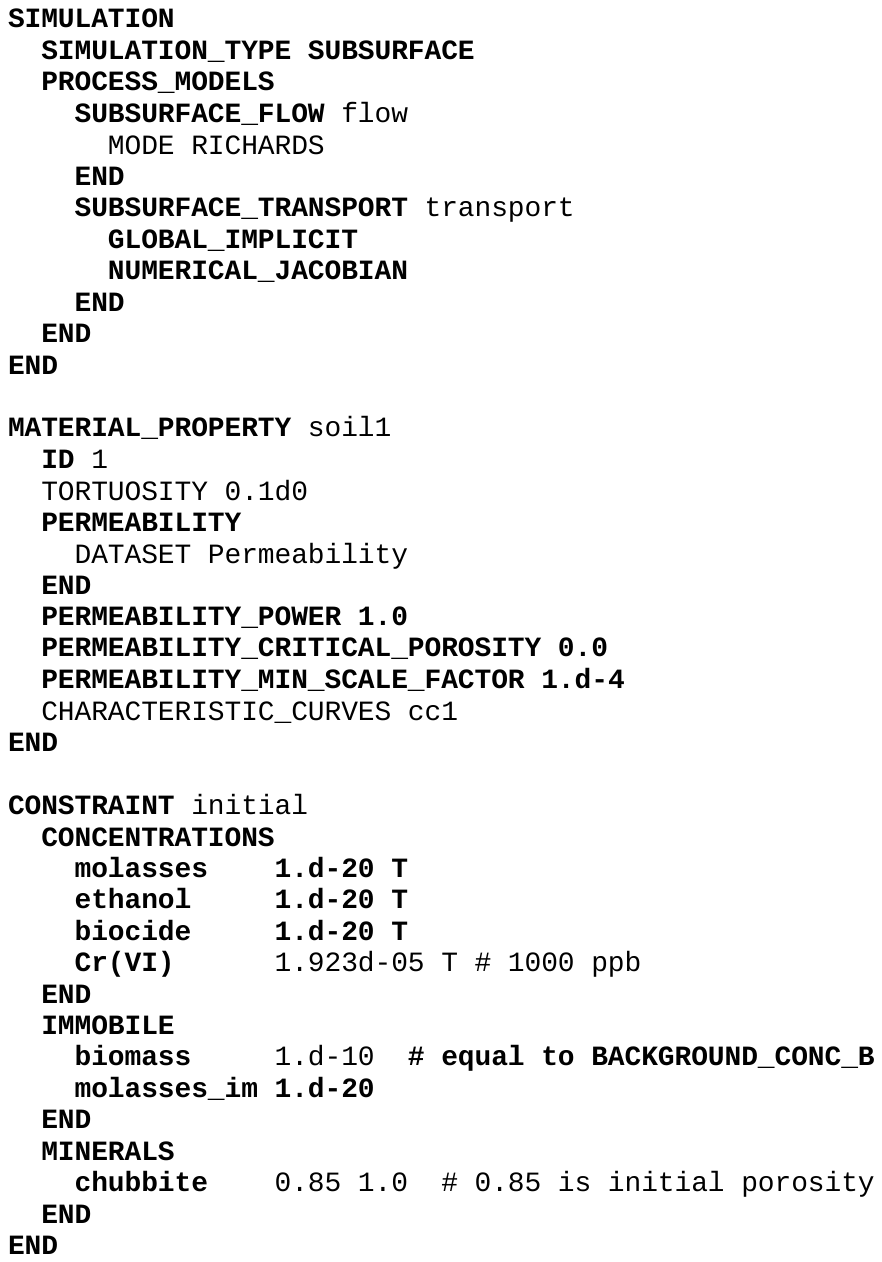}
\caption{Additional cards that require particular content in order for \textsc{chrotran} to work properly.
In the \texttt{SIMULATION} card, the \texttt{NUMERICAL\_JACOBIAN} option must be specified.
In the \texttt{MATERIAL\_PROPERTY} card, the OPTION \texttt{PERMEABILITY\_MIN\_SCALE\_FACTOR 1.d4} option should be set.
In \texttt{CONSTRAINT} cards, species that are not present should have small, but non-zero concentrations assigned.
The concentration of \texttt{NAME\_B} (\texttt{biomass}, here) should equal \texttt{BACKGROUND\_CONC\_B} in the \texttt{CHEMISTRY CARD}.
Finally, the initial porosity of the system is set by assigning the volume fraction of \texttt{NAME\_BIOMINERAL} (\texttt{chubbite}, here).
In general, bold text is required.
However, other options may be specified, if desired.}
\label{lst: simulation}
\end{figure}

A \textsc{chrotran} input file is of the same format as a \textsc{pflotran} input file.
Information on how to set up such a file is available in the \textsc{pflotran} user manual \citep{lichtner2015pflotran}.
However, to use \textsc{chrotran}'s additional functionality, a few of the input cards (top-level blocks, in \textsc{pflotran} jargon) must contain some particular content.
The required \texttt{CHEMISTRY} card format is shown in Figure \ref{lst: chemistry}, with bold text being mandatory and standard-weight text being user-alterable. 
The required \texttt{SIMULATION}, \texttt{MATERIAL\_PROPERTY}, and (initial) \texttt{CONSTRAINT} card formats are shown in Figure \ref{lst: simulation}, again with bold text being mandatory and standard-weight text being user-alterable.
Comments in the input file are preceded by the character \texttt{\#}.
\begin{table}
\caption{Relationship between the parameter names in the \texttt{CHEMISTRY} card (Figure \ref{lst: chemistry}) and the mathematical symbols shown in Section \ref{sec: equations}.}
$\:$\\
\centering
\begin{tabular}{c c l}
	\hline \hline
	Symbol & Units & Name in \texttt{CHEMISTRY} card \\
	\hline
	$\alpha$ & - &\texttt{EXPONENT\_B} \\
	\hline
	$B_\mathrm{min}$ & $\mathrm{mol\ L_b^{-1}}$&\texttt{BACKGROUND\_CONC\_B} \\
	\hline
	$\Gamma_{B}$ & $\mathrm{L\ mol^{-1}\ s^{-1}}$&\texttt{MASS\_ACTION\_B} \\
	$\Gamma_{CD}$ & $\mathrm{L\ mol^{-1}\ s^{-1}}$&\texttt{MASS\_ACTION\_CD} \\		 
	$\Gamma_{X}$ & $\mathrm{L\ mol^{-1}\ s^{-1}}$&\texttt{MASS\_ACTION\_X} \\
	\hline
	$K_B$ & $\mathrm{mol\ L_b^{-1}}$&\texttt{INHIBITION\_B} \\		 
	$K_{C}$ & $\mathrm{mol\ L^{-1}}$&\texttt{INHIBITION\_C} \\
	$K_D$ & $\mathrm{mol\ L^{-1}}$&\texttt{MONOD\_D} \\
	$K_I$ & $\mathrm{mol\ L^{-1}}$&\texttt{INHIBITION\_I} \\
	\hline
	$\lambda_{B_1}$ & $\mathrm{s^{-1}}$&\texttt{RATE\_B\_1} \\
	$\lambda_{B_2}$ & $\mathrm{s^{-1}}$&\texttt{RATE\_B\_2} \\
	$\lambda_C$ & $\mathrm{s^{-1}}$&\texttt{RATE\_C} \\
	$\lambda_{D}$ & $\mathrm{s^{-1}}$&\texttt{RATE\_D} \\
	$\lambda_{D_i}$ & $\mathrm{s^{-1}}$&\texttt{RATE\_D\_IMMOB} \\
	$\lambda_{D_m}$ & $\mathrm{s^{-1}}$&\texttt{RATE\_D\_MOBIL} \\
	\hline 
	$\rho_B$ & $\mathrm{mol\ L^{-1}}$ &\texttt{DENSITY\_B} \\
	\hline
	$S_C$ & - &\texttt{STOICHIOMETRIC\_C} \\
	$S_{D_1}$ & - &\texttt{STOICHIOMETRIC\_D\_1} \\
	$S_{D_2}$ & - &\texttt{STOICHIOMETRIC\_D\_2} \\
	\hline \hline
\end{tabular}
\label{tab: vartiable names}
\end{table}

\begin{figure}
\includegraphics[trim={2.5cm 21cm 10cm 2.5cm},clip]{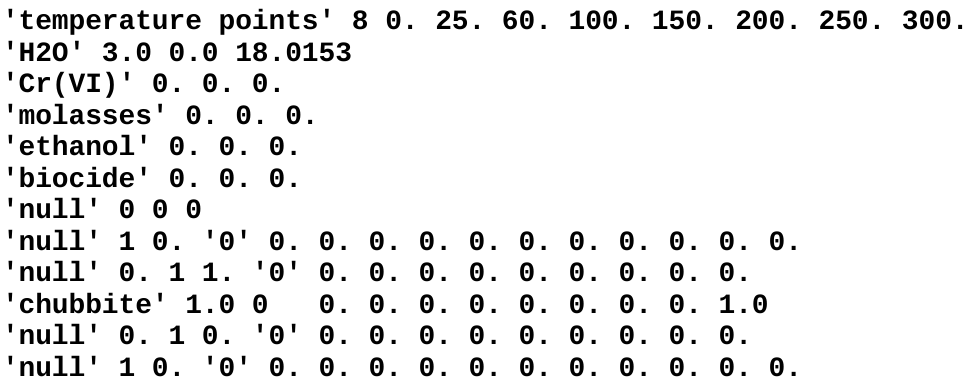}
\caption{Minimal \textsc{chrotran} chemistry database.
The text shown here should not be removed, however additional species may be added, if desired. See \textsc{pflotran} user manual for details on the database format.}
\label{lst: chem DB}
\end{figure}

In addition to these cards being properly formatted, there must exist a chemistry database at the (absolute or relative) path specified after the \texttt{DATABASE} keyword in the \texttt{CHEMISTRY} card, and it must, at a minimum contain the lines shown in Figure \ref{lst: chem DB}. The one exception to bold text being mandatory is that species \textit{names} can be changed at will, as long as there is consistency between the \texttt{CHEMISTRY} card and the chemistry database.
For instance, one could change all instances of the text \texttt{Cr(VI)} in both of those locations to \texttt{U(VI)} or all instances of the text \texttt{chubbite} to \texttt{etibbuhc}, with no alteration in execution behavior (besides, obviously, the species names used in the output files).

The chemistry database contains lines for five mobile species: water, plus the mobile species in the \textsc{chrotran} kinetics listed in Section \ref{sec: rt eqns}: $C$, $D_m$, $I$, and $X$. The database also contains a line for a ``dummy'' mineral species, \texttt{chubbite}, which does not correspond to any species previously mentioned. This species is treated as a mineral which is specified as inactive with respect precipitation/dissolution by setting its kinetic rate constant (\texttt{RATE\_CONSTANT}) to zero. The mineral is included as a surrogate for biomass and porous media volume in \textsc{chrotran} and is updated according to Equation \ref{eq: biovfupdate} to track $1-\theta(\bm{x},t)$.  
The initial volume fraction of \texttt{chubbite} thus defines the initial porosity.
The format of a chemistry database is discussed in more detail in the \textsc{pflotran} user manual.

Once you have saved your input file as, e.g. \texttt{test.in}, it is easy to run the code from the console.
Navigate to \texttt{<path of cloned repository>/src/pflotran}, and type \texttt{chrotran -pflotranin <path to input file>/test.in}.
The output of the simulation will be saved in the same directory as the input file.
Depending on the options specified in the input file, \textsc{chrotran} can save flow field velocities, concentrations of all species, permeabilities, and porosities at any specified times in an \texttt{.h5} format file.
This file format can be visualized natively using freely-available standalone tools such as VisIt and ParaView, and are also accessible from Python scripts by means of the \texttt{h5py} library and from Julia scripts by means of the \texttt{HDF5} package. 

\subsection{Quality assurance}
	The \textsc{pflotran} software from which \textsc{chrotran} derives its numerical flow and reactive transport solvers has gone through extensive quality assurance testing, has been benchmarked against other reactive transport solvers, and is used inside and outside the U.S. Department of Energy for mission-critical analytical work. The new bio-reactive transport model that is constitutive of \textsc{chrotran} is not available in any other software, so direct benchmarking is not possible. However, extensive quality testing has been performed by the developers. We have validated through batch and multidimensional simulations that \textsc{chrotran} does satisfy the governing equations we present for chemistry and permeability, and also that it gives plausible, physically consistent results for a wide range of scenarios.

\section{\textsc{chrotran} remediation case studies}
To demonstrate the capabilities of our software, we present two example studies, which together illustrate the interactions of all the types of chemical species it permits to be modeled, along with its treatment of bio-clogging. The input files for these two examples can be found in the \texttt{chrotran\_examples} directory of the \textsc{chrotran} repository.

\subsection{Remediation of Cr(VI) by molasses and ethanol co-injection}
\label{sec: application 1}
This study concerns the co-injection of molasses (electron donor, $D$) and ethanol (nonlethal bio-inhibitor, $I$) into a single well drilled in a heterogeneous aquifer with an appreciable background Cr(VI) concentration.
The competition between direct abiotic reduction of Cr(VI) by molasses and bio-reduction of Cr(VI), which exists since both reduction pathways consume the electron donor, along with the impact of suppressing the biomass growth is explored.
The basic parameters used are those shown in Figures \ref{lst: chemistry} and \ref{lst: simulation}, with changes as indicated below.

Four related simulations are performed on the same 100 $\times$ 100 m two-dimensional heterogeneous hydraulic conductivity field, with geometric mean conductivity $K_g = 10^{-4}\ \mathrm{m\ s^{-1}}$, a multi-Gaussian correlation structure with exponential semivariogram with correlation length of 4 m and $\sigma^2_{\ln K} = 2$.
Each simulation takes place over a span of 500 days and begins with $\epsilon=10^{-20}$ initial concentrations of all species, except $B(\bm{x},0)=B_{min}=10^{-10} \ \mathrm{mol\ L_b^{-1}}$ and $C=1.923\times10^{-5} \ \mathrm{mol\ L^{-1}}$ (1000 ppb Cr(VI)).
In all cases, there are no flow boundaries at the north and south of the domain ($y=0$ m and $y=100$ m), and constant head boundaries are imposed at the west and east of the domain ($x=0$ m and $x=100$ m) such that there is a drop of head of 0.28 m between these faces.
A single injection well exists at $(x,y)=(25\; \mathrm{m},50\; \mathrm{m})$.
For the first 10 days of the simulation, there is no injection into the well.
From 10 d to 30 d, injection is performed at the well with constant volumetric flow rate 272.55 $\mathrm{m^3 d^{-1}}$ with species concentrations discussed below.
From 30 d to 500 d, there is again no injection at the well.
A very large (arbitrary) $\rho_B$ is assumed, so as to eliminate the effect of biomass clogging from this simulation.

The four simulations differ in their chemistry only.
Two direct abiotic reduction rates are considered: $\Gamma_{CD} = 1\ \mathrm{L\ mol^{-1}\ s^{-1}}$ and $\Gamma_{CD} = 0\ \mathrm{L\ mol^{-1}\ s^{-1}}$, as are two different ethanol concentrations in the injection fluid: $I=1\ \mathrm{mol\ L^{-1}}$ and $I=\epsilon\ \mathrm{mol\ L^{-1}}$, in all four possible combinations.	
The injection fluid chemistry always has Cr(VI) concentration equal to the initial concentration ($C=1.923\times10^{-5} \ \mathrm{mol\ L^{-1}}$), ensuring that no chromium disappearance is due to dilution, and molasses concentration $D = 1\times10^{-2} \ \mathrm{mol\ L^{-1}}$.

\begin{figure}
\includegraphics[trim={1cm 1.5cm 0cm 1cm},clip]{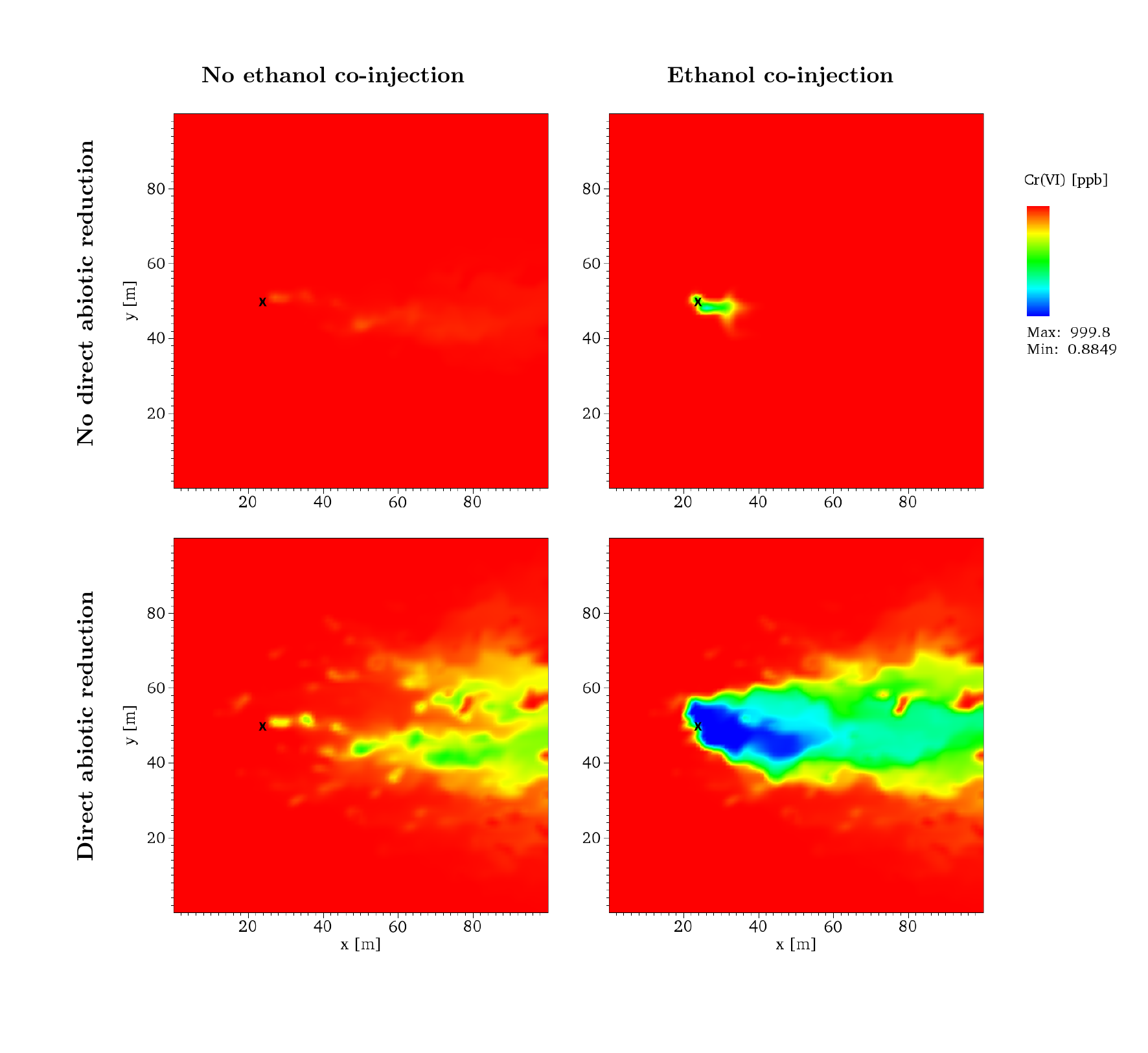}
\caption{Maps of Cr(VI) concentrations [ppb] in the aquifer 470 d after injection ceased in each of the four scenarios discussed in Section \ref{sec: application 1}.
Injection well location is denoted by a black \textit{X}.}
\label{fig: app 1}
\end{figure}

Concentrations of Cr(VI) for each scenario are shown in Figure \ref{fig: app 1}.
It is apparent that little persistent reduction due to biomass alone occurs, although ethanol co-injection does increase biomass footprint, which has a noticeable and persistent effect.
By contrast, the rapid abiotic reaction between Cr(VI) and a constituent of molasses has more impact.
This is attributable to the fact that molasses has a large reducing capacity, background concentrations of Cr(VI) are relatively low, and it has a retardation factor of around 150 (obtained from \cite{shashidhar2006}), meaning that it has the potential to form a persistent permeable reactive barrier around the well.
The better performance in the presence of ethanol is attributable to the fact that ethanol co-injection prevented consumption of molasses by the biomass during the injection phase, and so molasses persists over a larger area.

\subsection{Biomass clogging/unclogging due to acetate/dithionite injection}
\label{sec: application 2}
\textsc{chrotran} has the capability to model hydraulic conductivity reduction due to bio-fouling and the use of biocide as a remediation strategy.
To illustrate model capabilities, we perform a simulation of constant-head injection into a homogeneous aquifer in which the injection fluid amended initially with acetate ($D = 10^{-2}$ M) for the first 400 d, which is subsequently replaced with dithionite ($X = 3.5 $ M) for the remainder of the simulation.
The basic structure of the \textsc{chrotran} input file is the same as in the study outlined in Section \ref{sec: application 1} (this is to say, as shown in Figures \ref{lst: chemistry} and \ref{lst: simulation}), but with different \textsc{chrotran} parameter values, as shown in Table \ref{tab: ex 2 params}. We here make the reasonable \citep[p. 361]{ritmann2004}
assumption that biomass has the same density as water (recall that we everywhere use the interpretation that 1 mol of biomass is defined as 1 g of biomass). 

The simulation is performed on a 50 m square homogeneous hydraulic conductivity field, with constant hydraulic conductivity $K = 9.8\times 10^{-5}\ \mathrm{m\ s^{-1}}$. Each simulation takes place over a span of 500 days, and begins with $\epsilon=10^{-20}$ initial concentrations of all species, except $B(\bm{x},0)=B_{min}=10^{-10} \ \mathrm{mol\ L_b^{-1}}$ and $C=1.923\times10^{-5} \ \mathrm{mol\ L^{-1}}$.
No flow boundaries are imposed at the north and south of the domain ($y=0$ m and $y=50$ m), and constant head boundaries are imposed at the west and east of the domain (head 0.28 m at $x=0$ m and head 0 m at $x=50$ m).
A single injection well exists at $(x,y)=(25\;\mathrm{m},25\;\mathrm{m})$, and constant head of 0.28 m is imposed at its location.

\begin{figure}
\centerline{
\includegraphics[trim={1cm 7cm 2cm 1cm},clip]{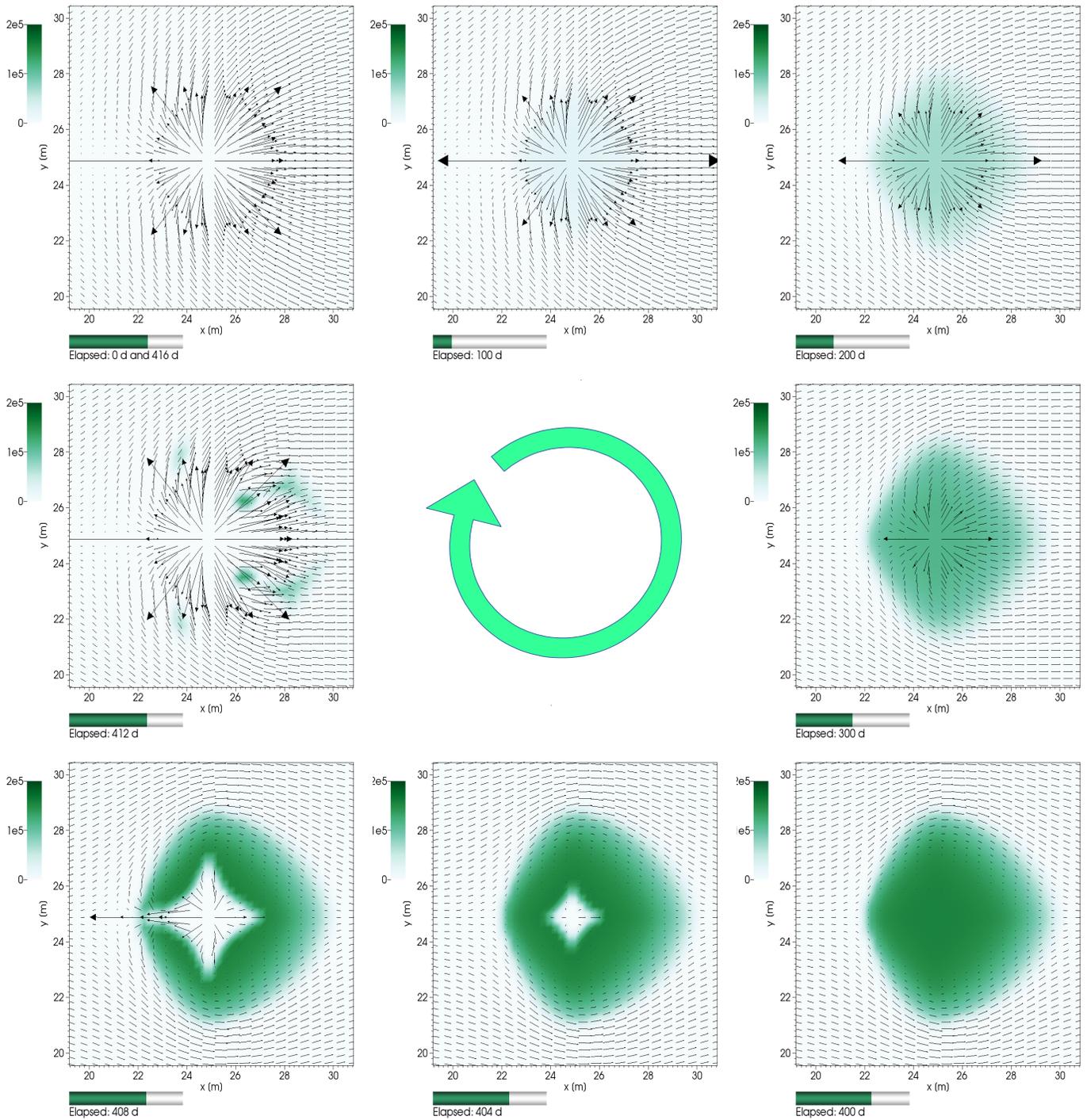}
}
\caption{A sequence of snapshots of the cell-center groundwater seepage velocity fields and biomass concentration distributions in the example of Section \ref{sec: application 2}.
Velocity magnitude is indicated by arrow length and direction by arrow orientation; the arrow tails are located at the cell center; biomass concentration [$\mathrm{g\ m^{-3}}$] is indicated by green intensity in each superimposed map.
The initial condition snapshot is shown in the upper left corner, with time increasing in the clockwise direction, until the initial condition is reached again at 416 d.
The same scale is used in each snapshot.}
\label{fig: app 2}
\end{figure}

A sequence of quiver plots representing the velocity field at nine points in time, superimposed on the intensity of biomass concentration are shown in Figure \ref{fig: app 2}.
During the first 400 d of the simulation, biomass concentration grows in the vicinity of the well, until hydraulic conductivity drops to zero at the well until no influx occurs there; only ambient flow is apparent, flowing around the impermeable barrier near the well.
At this point, the biomass has become useless for bioremediation, as contaminated aquifer water no longer travels through it.
However, at 400 d, dithionite is introduced into the injection fluid and effectively
 eliminates biomass in the vicinity of the well.
The region containing dithionite is relatively sterile and grows outwards until the biomass concentration approaches background, and the initial flow regime is recovered at 416 d.
Because initial and final conditions are the same, this cycle may be performed indefinitely. 

\begin{table}
\caption{\textsc{chrotran} parameter values used in the bio-fouling example in Section \ref{sec: application 2}.}
$\:$\\
\centering
\begin{tabular}{c c l}
	\hline \hline
	Symbol & Value & Units\\
	\hline
	$\alpha$ & 1 & - \\
	\hline
	$B_\mathrm{min}$ & $10^{-10}$ &$\mathrm{mol\ L_{b}^{-1}}$\\ 
	\hline
	$\Gamma_{B}$& $2.6 \times 10^{-2}$ &$\mathrm{L\ mol^{-1}\ s^{-1}}$\\
	$\Gamma_{CD}$& 1 &$\mathrm{L\ mol^{-1}\ s^{-1}}$\\
	$\Gamma_{X}$& $2.6 \times 10^{-5}$ &$\mathrm{L\ mol^{-1}\ s^{-1}}$\\
	\hline
	$\lambda_{B_1}$& $10^{-5}$ &$\mathrm{s^{-1}}$\\
	$\lambda_{B_2}$& $10^{-15}$ &$\mathrm{s^{-1}}$\\
	$\lambda_{C}$& $10^{-10}$ &$\mathrm{s^{-1}}$\\
	$\lambda_{D}$& 0 &$\mathrm{s^{-1}}$\\
	$\lambda_{D_i}$& 1.5 &$\mathrm{s^{-1}}$\\
	$\lambda_{D_m}$& $10^{-2}$ &$\mathrm{s^{-1}}$\\
	\hline
	$K_B$& $5 \times 10^{2}$ & $\mathrm{mol\ L_{b}^{-1}}$\\
	$K_C$& $10^{-7}$ & M\\
	$K_D$& $10^{-6}$ & M\\
	$K_I$& 1 & M\\
	\hline
	$\rho_B$& $10^{3}$ & $\mathrm{mol\ L^{-1}}$\\
	\hline
	$S_C$& $3.3\times 10^{-1}$ & - \\
	$S_{D_1}$& $10^{-5}$ & - \\
	$S_{D_2}$& $2.0833\times 10^{-2}$ & - \\
	\hline \hline
\end{tabular}
\label{tab: ex 2 params}
\end{table}

\section{Summary and conclusions}
For modeling in situ remediation of aqueous groundwater contaminants by injection of aqueous amendment, we recognized the importance of mathematical formulations and numerical codes that can represent three-dimensional fluid flow and multi-species contaminant transport in heterogeneous aquifers with arbitrary injection regimes.
For the particularly important case of heavy metal remediation, a number of contaminant-remediation processes (pathways) are susceptible to a unified modeling framework: bio-reduction, bio-precipitation, and direct reduction by the chemical amendment.
There have previously existed no general tools appropriate for modeling such interventions.
With this background in mind, we developed a mathematical model that describes the reactive transport dynamics of an amendment (containing any combination of electron donor, non-lethal bio-inhibitor, and biocide) with biomass and aqueous heavy metal contaminant.
We also implemented the mathematical model in a novel computational framework, called \textsc{chrotran}, that is based on the open-source code \textsc{pflotran}. \textsc{pflotran}'s modularity and the reaction sandbox capability allowed us to implement the model easily without making any changes to the flow and transport part of \textsc{pflotran}.
\textsc{chrotran} can harness the high-performance computing capabilities of \textsc{pflotran} which allows for simulations of complex models with large number of computational cells and degrees of freedom.
We described our computer implementation and explained how to use \textsc{chrotran} to solve practical problems.

We also considered two demonstration studies related to chromium remediation.
The presented synthetic problems were formulated to be consistent with real-world groundwater contamination problems and 
illustrate the capability of \textsc{chrotran} to aid in the engineering design process.
In one of the studies, we discovered that, contrary to much existing theory, Cr(VI) reduction was maximized by injecting molasses and suppressing biomass growth to maximize the direct, abiotic reduction reaction.
In the other, we showed the feasibility of pulsed injection of bio-stimulant and biocide to alleviate bio-fouling in the context of ongoing bioremediation.

We observe that because of the abstraction of our model and its parametric flexibility, the \textsc{chrotran} equations can be used to model other reactive transport behaviors besides the heavy metal bio-reduction that we have focused upon, including basic advection-dispersion-reaction interaction (between $C$ and $D$, in the absence of $B$).
The bio-reduction model captures any biodegradation that can be represented using a Monod equation, as long as the contaminant represented by $C$ is non-sorbing, and it does not explicitly require the contaminant to be reduced. This potentially allows for modeling the biodegradation of a wide range of organic contaminants, which include but are not limited to hydrocarbons, chlorinated solvents, pesticides, and volatile organic compounds.

\section*{Acknowledgments}
The authors acknowledge the support of the LANL Environmental Programs.

\end{document}